# Inhibition of tunnelling and edge state control in polariton topological insulators

Yiqi Zhang,[1] Yaroslav V. Kartashov,[2,3,*] Yanpeng Zhang,[1] Lluis Torner[2,4] and Dmitry V. Skryabin[5,6]

[1]Key Laboratory for Physical Electronics and Devices of the Ministry of Education & Shaanxi Key Lab of Information Photonic Technique, Xi'an Jiaotong University, Xi'an 710049, China
[2]ICFO-Institut de Ciencies Fotoniques, The Barcelona Institute of Science and Technology, 08860 Castelldefels (Barcelona), Spain
[3]Institute of Spectroscopy, Russian Academy of Sciences, Troitsk, Moscow, 108840, Russia
[4]Universitat Politecnica de Catalunya, 08034, Barcelona, Spain
[5]Department of Physics, University of Bath, Bath BA2 7AY, United Kingdom
[6]ITMO University, Kronversky Avenue 49, St. Petersburg 197101, Russia

**Abstract:** We address the inhibition of tunnelling in polariton condensates confined in a potential landscape created by a honeycomb array of microcavity pillars in the presence of spin-orbit coupling and Zeeman splitting in the external magnetic field. The coupling rate between the microcavity pillars can be strongly impacted even by weak out-of-phase temporal modulations of the depths of the corresponding potential wells. When such a modulation is implemented in truncated honeycomb arrays that realize a polariton topological insulator, which supports unidirectional edge states in the presence of spin-orbit coupling and Zeeman splitting, it allows controlling the velocity of the states. The origin of the phenomenon is the dynamical modulation with a proper frequency, which notably changes the dispersion of the system and the group velocity of edge states. We show that such a control is possible for modulation frequencies close to resonances for inhibition of tunnelling in a two-well configuration. Edge states considerably slow down, and even stop completely, when the modulation frequency approaches a resonant value, while above such frequency splitting of the edge states into wavepackets moving with different velocities occurs.

     Topological insulators are a topic of highest interest currently. One of their most representative features, which distinguishes them from conventional insulators, is the existence of unidirectional in-gap edge states that emerge when the topological insulator is placed in contact with a material of distinct topology[1,2]. Due to their topological nature and absence of states into which scattering can occur, edge states survive in the presence of disorder and can bypass structural defects. Topological insulators, first studied in the context of the quantum Hall effect[1-4], represent nowadays an interdisciplinary concept that has been demonstrated in diverse areas of physics, including acoustics[5,6], mechanics[7], physics of cold atoms[8-11], and photonics[12-35], to name a few. Demonstrations in photonic systems include gyromagnetic photonic crystals[12,13], semiconductor quantum wells[14], arrays of coupled resonators[15,16], metamaterial superlattices[17] and periodic structures[18,19], helical waveguide arrays[20–23], parity-time symmetric systems[24], polaritons in microcavities[25-31], as well as various dissipative structures allowing lasing in topological states[32-35]. Combination of topological and nonlinear material properties enable the formation of unidirectional topological quasi-solitons, as predicted in waveguide arrays[36-38] and polaritonic systems[29,30,39,40].

     The uncommon dispersion of the edge states, whose unidirectional motion becomes apparent when wavepackets with localized envelope are used for edge state excitation, is determined by the physical factors leading to opening of the topological gap and, most importantly, by the internal structure (symmetry) of the insulator. Thus, photonic Floquet[20-22] and polariton[25,29] insulators are most frequently constructed using honeycomb arrays of waveguides or microcavity pillars, respectively. The symmetry of the underlying array determines the momentum intervals where edge states can appear for a particular truncation type. Elucidation of new mechanisms that allow controlling the dispersion of the topological edge states, hence the transport properties of the system, is an important problem from both fundamental and practical points of view.

Here we expose one such mechanism that keeps the internal structure of the topological insulator unchanged and that is based on shallow periodic temporal modulations of the depths of the potential wells in the honeycomb array. It relies on the phenomenon of inhibition of tunnelling that was introduced in quantum mechanics and that leads to a considerable suppression of tunnelling for certain modulation frequencies in time-modulated potentials[41,42]. Nowadays, the control of tunnelling by periodic forces is routinely used in non-topological systems[43,44]. In photonics, this phenomenon has been observed in waveguide arrays modulated in the direction of light propagation[45,46] (see also the recent review[47]). Honeycomb structures frequently used for construction of topological insulators are especially suitable for observation of inhibition of tunnelling, since this effect requires out-of-phase modulation of the depths of all neighbouring potential wells[48], a property that is easily realizable in honeycomb array.

We consider a polariton topological insulator built as a honeycomb array of microcavity pillars that can be readily fabricated experimentally and that has been already used for demonstration of nontopological polaritonic edge states[49,50]. Existence of topological edge states in such a system relies on the combination of polarization-dependent tunnelling between neighbouring pillars, analogous to spin-orbit coupling effect[51,52], and Zeeman splitting in a magnetic field through the excitonic component of the polariton condensate. We show that the group velocity of the edge states can be controlled by a weak periodic out-of-phase temporal modulation of the neighbouring microcavity pillars, which can be realized, e.g., using electro-optic[53], acoustic[54], free-carrier based[55], and other[56,57] techniques. When the modulation frequency approaches one of the resonant values, the group velocity of the edge states can be decreased practically to zero, thus leading to a decreasing displacement along the edge for states with a localized envelope. The phenomenon takes places with a suppressed emission into the bulk due to the topological protection, which ensures motion of the wavepacket only along the edge with a velocity controlled by the modulation.

The evolution of a spinor wavefunction $\boldsymbol{\Psi}=(\psi_+,\psi_-)^{\mathrm{T}}$ describing a polariton condensate in a time-modulated and spatially-truncated honeycomb array of microcavity pillars is governed by the system of the dimensionless coupled Schrödinger equations[25,29]:

$$i\frac{\partial \psi_\pm}{\partial t} = -\frac{1}{2}\left(\frac{\partial^2}{\partial x^2}+\frac{\partial^2}{\partial y^2}\right)\psi_\pm + \beta\left(\frac{\partial}{\partial x}\mp i\frac{\partial}{\partial y}\right)^2 \psi_\mp + \mathcal{R}(x,y,t)\psi_\pm \pm \Omega\psi_\pm. \qquad (1)$$

We adopt a circular polarization basis, where the relation between the wavefunctions of the spin-positive and spin-negative components and wavefunctions of conventional TE (subscript y) and TM (subscript x) polarizations reads as $\psi_\pm=(\psi_{\mathrm{x}}\mp i\psi_{\mathrm{y}})/2^{1/2}$; the parameter $2\beta=(m_{\mathrm{x}}-m_{\mathrm{y}})/(m_{\mathrm{x}}+m_{\mathrm{y}})$ accounts for spin-orbit coupling[51,52]; $m_{\mathrm{x,y}}$ are the effective masses of the TM and TE polaritons, respectively; the term $\sim\Omega$ is the Zeeman energy splitting in the external magnetic field; we also assume linear small-density regime. Time-dependent potential landscape produced by microcavity pillars is described by the function $\mathcal{R}=[1+\mu\sin(\omega t)]\mathcal{R}_{\mathrm{A}}(x,y)+[1-\mu\sin(\omega t)]\mathcal{R}_{\mathrm{B}}(x,y)$, where $\mathcal{R}_{\mathrm{A}}$ and $\mathcal{R}_{\mathrm{B}}$ describe two standard sublattices[36] of the honeycomb structure, that exhibit weak ($\mu\ll 1$) out-of-phase modulation with frequency $\omega$. Each sublattice $\mathcal{R}_{\mathrm{A,B}}=-p\sum_{n,m}\mathcal{Q}(x-x_n,y-y_m)$ is composed from Gaussian potential wells $\mathcal{Q}=\exp[-(x^2+y^2)/d^2]$ with the width $d$ and depth $p$, the separation between neighbouring potential wells in the resulting array $\mathcal{R}$ is given by $a$. Each potential well in $\mathcal{R}$ oscillates in time out-of-phase with its three nearest neighbours, which is a crucial ingredient for tunnelling inhibition. Here we assume that the array is truncated along the $x$-axis and that it is periodic along the $y$-axis, so that two zigzag edges appear for $-\mathcal{R}$ profile, as depicted in Fig 1(a). Results that we obtained for the case of bearded edges are qualitative similar, thus they are not shown here. The $y$ period of the structure is $T=3^{1/2}a$. In Eq. (1) all spatial coordinates are normalized to the characteristic length $L$, all energy parameters (including potential depth and Zeeman splitting) are normalized to the characteristic energy $\epsilon_0=\hbar^2/mL^2$, where $m=2m_{\mathrm{x}}m_{\mathrm{y}}/(m_{\mathrm{x}}+m_{\mathrm{y}})$ is the effective polariton mass, while time is normalized to $\hbar\epsilon_0^{-1}$. For the calculations, we set $a=1.5$, $d=0.5$, $p=8$, which for a characteristic length of $L=1\,\mu\mathrm{m}$ corresponds to $1\,\mu\mathrm{m}$-wide potential wells with cen-tre-to-centre separation of $1.4~\mu$m and depths of $5.6$ meV. In these estimates the effective polariton mass $m \sim 10^{-34}$ kg was used, which gives $\epsilon_0 \sim 0.7$ meV.

Topological edge states appear due to simultaneous action of spin-orbit coupling and Zeeman splitting. These two effects, acting together, break the time-reversal invariance in the Eq. (1) that, according to work [3], should lead to opening of the topological gap between former Dirac points in the spectrum of the honeycomb array[25,29]. To illustrate the effect for the truncated array from Fig. 1(a), we first switch off the temporal modulation by setting $\mu=0$ and search for eigenmodes of the structure in the form of Bloch waves $\psi_\pm(x,y,t) = u_\pm(x,y)\exp(iky+i\epsilon t)$, which are periodic $u_\pm(x,y) = u_\pm(x,y+T)$ along the $y$-axis and localized $u_\pm(x \to \pm\infty, y) = 0$ along the $x$-axis. Such Bloch modes are characterized by the dependence of their energy $\epsilon$ on Bloch momentum $k$, which should be periodic with a period $K = 2\pi/T$ according to the Floquet theorem. Such dependence is found from the linear eigenvalue problem:

$$\epsilon u_\pm = (1/2)[\partial^2/\partial x^2 + (\partial/\partial y + ik)^2]u_\pm - \mathcal{R}(x,y)u_\pm - \beta[\partial/\partial x \mp i(\partial/\partial y + ik)]^2 u_\mp \mp \Omega u_\pm, \quad (2)$$

and presented in Fig. 1(b), which shows a zoom of the dispersion relation between Dirac points. We studied a structure containing 44 microcavity pillars per unit cell (one $y$-period of the array): four such periods are shown in Fig. 1(a). The presence of the topological gap with unidirectional edge states in it (red and blue lines between remnants of Dirac points at $k=K/3$ and $k=2K/3$) is readily visible. Since our system is spinor and transforms at $\beta=0$ into two independent equations, there are always two similar groups of bands shifted by $2\Omega$ and characterized by the dominance of different components, $\psi_+$ or $\psi_-$, in spinor wavefunction[29]. We deliberately selected a sufficiently large Zeeman splitting $\Omega=2.35$ to have two upper bands from one of these groups well isolated from the rest of the spectrum [Fig. 1(b)]. The wavefunctions emerging from these bands have a dominating $\psi_-$ component. The width of the topological gap generally increases with increase of the spin-orbit coupling strength $\beta$. From now on we set $\beta=0.3$.

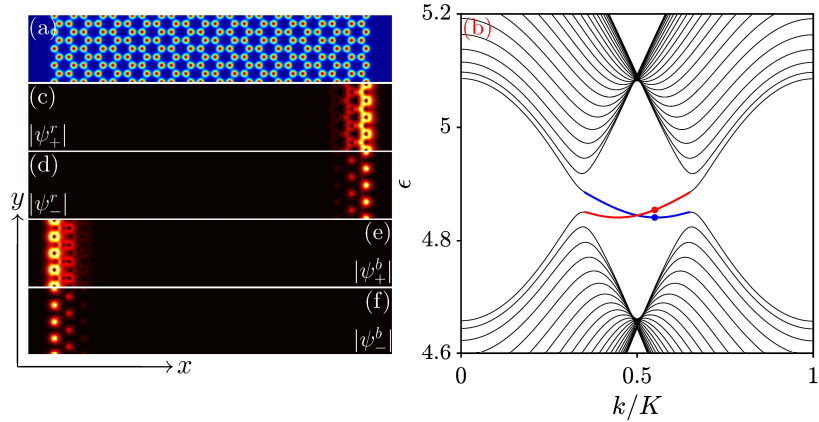

Fig. 1. (a) Potential created by an array of microcavity pillars with zigzag-zigzag edges. (b) Band structure corresponding to an array with 44 pillars per unit cell. Red and blue curves are associated with the edge states; dots are located at $k=0.55$K. (c)-(f) Spatial distributions $|\psi_\pm|$ corresponding to dots in (b). The domain shown in (a),(c)-(f) is $x \in [-28, +28]$ and $y \in [-2T, +2T]$.

The topological mode associated to the red branch (superscript $r$) resides on the right edge of the insulator, with a group velocity $v = -\partial\epsilon/\partial k$ at $k=0.55$K that is negative. Therefore, the mode moves along the negative direction of the $y$-axis. In contrast, the mode from the blue branch (superscript $b$) resides on the left edge of the insulator and at $k=0.55$K moves in the positive direction of the $y$-axis. One observes the appearance of charge-2 vortices in each well in the $\psi_+$ component, a phenomenon that is a consequence of the spin-orbit coupling specific for polaritons.

As the group velocity of the edge states is defined by $\partial\epsilon/\partial k$, their dispersion is defined by $\partial^2\epsilon/\partial k^2$. The motion of the edge state becomes apparent in the presence of a broad localized envelope along the $y$-axis. An illustrative example of the evolution of the edge state from the right edge [red dot in Fig. 1(b)] with the Gaussian envelope $\exp[-(y-y_{\text{in}})^2/w_{\text{in}}^2]$ with $y_{\text{in}}=20$ and $w_{\text{in}}=6T$ in the unmodulated array is shown in Fig. 2. According to Fig. 1(b) the velocity of such state is $v\approx-0.1$, therefore it traverses about 60 periods of the structure and notably broadens over the time interval shown. Notice that due to topological protection, the wavepacket moves along the edge and does not penetrate into the bulk.

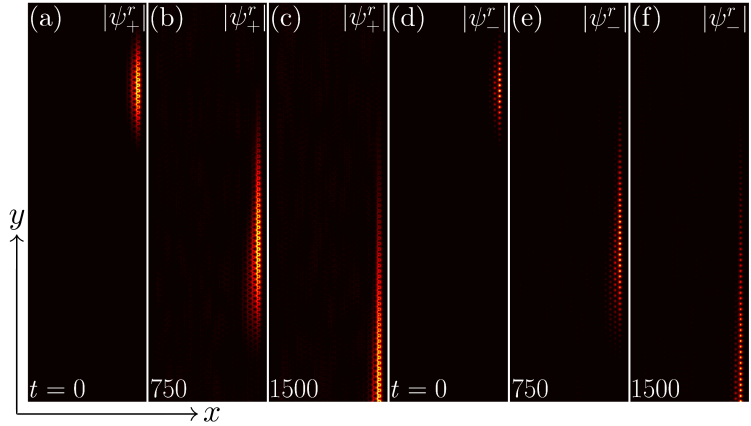

Fig. 2. Snapshots illustrating the evolution of topological edge states with an envelope in the unmodulated array. (a)-(c) Spin-positive component, (d)-(f) spin-negative component. The $|\psi_\pm|$ distributions are shown within $-28\leq x\leq 28$ and $-130\leq y\leq 60$ windows.

Our main goal here is showing that the velocity of the wavepacket along the edge can be efficiently controlled, while its broadening can be practically cancelled, by a suitable weak temporal modulation of the potential (microcavity pillars). To gain insight into the physics underlying the mechanism, we first address the impact on tunnelling of the out-of-phase modulation of potential wells in a simple two-pillar structure. Note that we take into account all physical effects that are included into model (1), i.e. Zeeman splitting and spin-orbit coupling. The latter dictates that the mode of a single potential well exhibits a bell-shaped profile in the $\psi_-$ component and carries a charge-2 vortex in the substantially less localized $\psi_+$ component, as visible in the insets of Fig. 3(a). When the pillars are brought in contact with each other one observes Josephson oscillations, namely periodic switching between the pillars. For selected parameters of the structure (that we optimized to get sufficiently large velocity $v$ of the edge states), the period of oscillations at $\mu=0$ is $T_\text{b}\approx 13.15$. If an out-of-phase temporal modulation of the potential wells is imposed, the evolution of the $t$-periodic system is governed by quasi-energies, rather than by usual energy $\epsilon$ as in the static system. The solutions in this case can be found in the form $\psi_\pm(t)=u_\pm(t)\exp(i\epsilon t)$, where $\epsilon$ is the quasi-energy defined modulo $\omega$, and $u_\pm(t)=u_\pm(t+2\pi/\omega)$ is the time-periodic function whose period is defined by the period of the temporal modulation. As we found numerically, in modulated two-pillar system quasi-energies $\epsilon$ of localized modes practically (but not exactly) coincide at a set of resonance frequencies $\omega$ that depend on the system parameters, such as depths of potential wells, separation between them, and spin-orbit coupling strength. At these frequencies the modes of the system do not experience dephasing[41,45,46], and therefore the input wavepacket that may excite an arbitrary combination of such modes remains practically unchanged upon evolution, i.e., tunnelling is practically arrested. For the parameters studied in this paper, the largest modulation frequency at which quasi-energies coincide in two-pillar system is $\omega\approx 0.75$.

To illustrate the above-described effect using a single-pillar excitation, we calculate the time-averaged norm in the excited (right) potential well defined as:

$$U_{\text{av}} = \frac{1}{4T_{\text{b}}U_0} \int_0^{4T_{\text{b}}} dt \int_{-\infty}^{+\infty} dy \int_0^{+\infty} \mathbf{\Psi}^\dagger(x,y,t)\mathbf{\Psi}(x,y,t) dx, \qquad (3)$$

where $U_0$ is the norm in the excited well in the initial moment of time $t=0$, and averaging is performed over $t=4T_{\text{b}}$. The dependence $U_{\text{av}}(\omega)$ is shown in Fig. 3(a). Maxima in this dependence are indicators of inhibited tunnelling – as one can see, there are multiple resonances, each of which corresponds to a minimum of the difference between quasi-energies. Inhibition is not complete, because in the system under study the quasi-energies do not coincide exactly (otherwise one would have $U_{\text{av}} = 1$ at the maxima), but coupling is delayed drastically in resonances. The phenomenon is visible in the evolution dynamics in $y=0$ cross-sections for $\omega=0.65$ [out of resonance, blue dot in Fig. 3(a)] and for $\omega=0.75$ [primary resonance, red dot in Fig. 3(a)], in Figs. 3(b) and 3(c), respectively. At resonance [Fig. 3(c)] the excitation remains in the excited well at all times and the fraction of norm switching to neighbouring well is negligible. The same picture is observed in secondary resonances, but quality of the tunnelling inhibition reduces with the decrease of the resonant frequency. In Fig. 3 we use optimal modulation depth $\mu=0.2$, because too shallow modulations cannot inhibit coupling, while too deep modulations result in strong radiation. We observed that resonance frequencies grow almost linearly with $\mu$. Notice that inhibition of tunnelling persists even if the presence of a phase shift between the modulations of depths of the two potential wells. However, its efficiency is maximal when the modulation is out-of-phase. This is the case for both, the two-pillar system and the extended topological insulator.

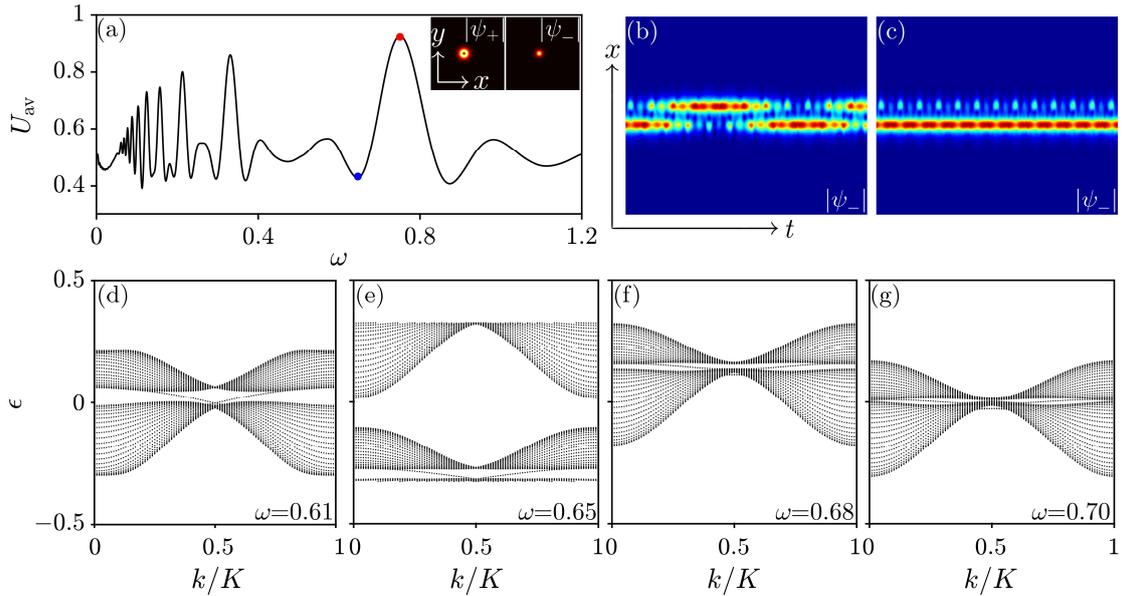

Fig. 3. (a) Time-averaged norm $U_{\text{av}}$ versus modulation frequency $\omega$ in a two-pillar system. Insets show the $\psi_+$ and $\psi_-$ components in a mode of single pillar for $-8 \leq x, y \leq 8$. Evolution dynamics ($\psi_-$ component in $y=0$ cross-section) in a modulated two-pillar system at (b) $\omega=0.65$ [blue dot in (a)], and (c) $\omega=0.75$ [red dot in (a)]. In (b),(c) we show dynamics within windows $0 \leq t \leq 8T_{\text{b}}$ and $-8 \leq y \leq 8$. (d)-(g) Quasi-energy bands $\epsilon(k)$ of the modulated topological insulator for $\mu=0.2$ and different modulation frequencies indicated in each panel. We use the same vertical scale in panels (d)-(g).

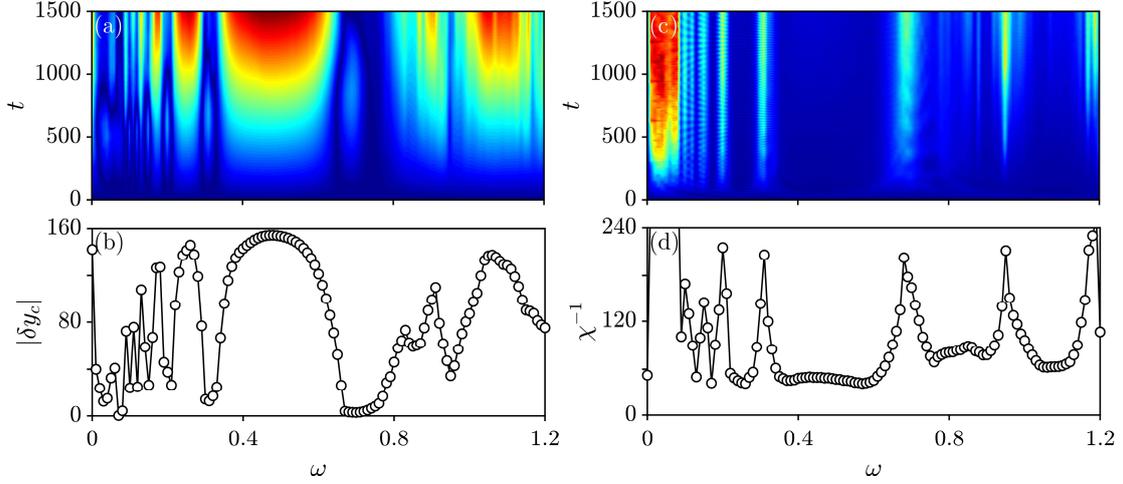

Fig. 4. (a) Displacement $|\delta y_c|$ of the edge state along the edge in topological insulator versus evolution time $t$ and modulation frequency $\omega$. (b) Displacement $|\delta y_c|$ as a function of $\omega$ at $t=1500$. The inverse form-factor $\chi^{-1}$ of the edge state versus modulation frequency and evolution time (c) and versus modulation frequency at $t=1500$ (d). In all cases $\mu=0.2$.

Since tunnelling between potential wells determines the evolution of the excitations also in complex extended systems, one expects that a similar resonant phenomenon may occur in the topological insulator in the presence of out-of-phase modulation of the potential wells. However, importantly, in that case the modes of the system are extended edge states having nonzero transverse momentum $k$. To study the impact of the potential modulation on their evolution we use the same input conditions as in Fig. 2 (edge state with broad envelope), vary the modulation frequency $\omega$ and follow the evolution of the centre of mass of the wavepacket given by the expression:

$$y_c(t) = \iint y\mathbf{\Psi}^\dagger\mathbf{\Psi}dxdy \Big/ \iint \mathbf{\Psi}^\dagger\mathbf{\Psi}dxdy. \qquad (4)$$

Our calculations reveal a strongly nonmonotonic dependence of the wavepacket centre displacement $\delta y_c = y_c - y_{\text{in}}$ on the modulation frequency [Fig. 4(a)], which becomes apparent already for relatively short evolution times, in the range $t \sim 100$. The typical dependence $|\delta y_c(\omega)|$ at fixed time $t=1500$ is displayed in Fig. 4(b). There is a clear correlation between the frequencies at which the displacement is minimal (practically zero around $\omega = 0.75$) and the resonances for tunnelling inhibition in Fig. 3(a). For frequencies between the resonant ones the edge states traverse considerable distances ($\sim 60$ periods) along the edge. The displacement is practically zero in the principal (i.e. right) resonance and in the second (next to principal) resonance. It is also relevant to show the expansion rate of the edge states during evolution, and to quantify it we introduced the form-factor, defined as:[58]

$$\chi^2 = \iint (\mathbf{\Psi}^\dagger\mathbf{\Psi})^2 dxdy \Big/ \left(\iint \mathbf{\Psi}^\dagger\mathbf{\Psi}dxdy\right)^2. \qquad (5)$$

The quantity $\chi^{-1}$ characterizes the width of the wavepacket. In Figs. 4(c) and 4(d), the dependence $\chi^{-1}(\omega)$ is shown during evolution and in the final moment of time $t=1500$, respectively. There is a clear correlation between the width and the displacement. For example, the minimal width is achieved close to the resonance frequencies, where displacement is also minimal.

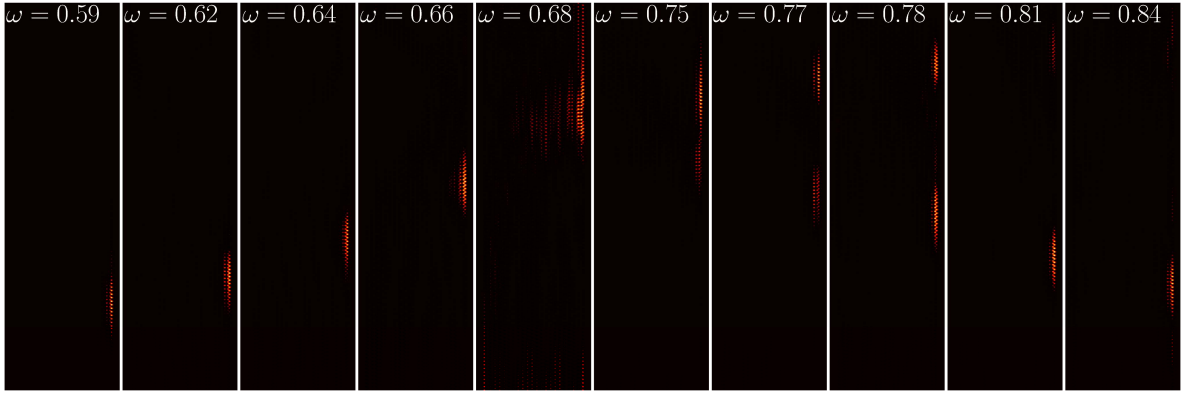

Fig. 5. Output $|\psi_-|$ distributions at $t=1500$ for modulation frequencies around the principal resonance. Frequency values are indicated in each panel. In all cases $\mu=0.2$, $k=0.55\,\text{K}$.

Figure 5 shows the output $|\psi_-|$ distributions at $t=1500$ for modulation frequencies around the principal resonance. Remarkably, already for frequencies $\omega\approx 0.59$ the broadening of the wavepacket is notably reduced in comparison with the case of the unmodulated topological insulators. The displacement along the edge monotonically decreases (the interval of variation of wavepacket position is enormous and can be further increased for longer evolution times) when the modulation frequency approaches the primary resonance ($\omega=0.59-0.68$). This constitutes the central result of our paper, i.e., the possibility to control displacements in the topological system by a weak modulation of the structure. To gain further insight, we calculated the quasi-energy bands of the modulated topological insulator for different modulation frequencies $\omega$. Examples of quasi-energy bands are presented in the bottom row of Fig. 3. In the $y$-periodic topological insulator, quasi-energies $\epsilon$ are functions of the Bloch momentum $k$. One can see that the topological gap with edge states in it persists even for the modulation amplitude $\mu=0.2$ used here. However, the width of the gap notably decreases when the modulation frequency approaches the resonant values. Because the width of the gap decreases, the derivatives $\partial\epsilon/\partial k$ and $\partial^2\epsilon/\partial k^2$ for topological branches determining the velocity and effective dispersion of the corresponding topological edge states decrease too, hence the reduced displacement and dispersion close to resonant frequencies. Figure 3(g) shows that the topological gap shrinks at $\omega\approx 0.70$ (it should be stressed that in periodic system the resonant frequency may depart from resonant frequency of two-pillar system). This is consistent with a strong expansion of the wavepacket into the bulk of the array observed in direct simulations shown in Fig. 5. Thus, in resonances the topological properties of the system degrade and as a result in a narrow range of frequencies one observes radiation into the bulk. When modulation frequency increases above the resonant value, the wavepacket splits into two wavepackets that may move in opposite directions, while radiation into the bulk cancels ($\omega=0.78$). The amplitude of one of these wavepackets gradually decreases and for high frequencies one again observes considerable shifts in the negative direction of the $y$-axis ($\omega=0.84$).

Similar dynamics was observed for other transverse momenta $k$ and not only around principal, but also around secondary resonances in $\omega$. Figure 6 shows transformation of the output state with increase of modulation frequency, when it approaches the second resonance. Just as in the primary resonance, the displacement of the wavepacket monotonically decreases with $\omega$, until the wavepacket experiences an abrupt destruction around the resonance at $\omega=0.304$.

Summarizing, we studied the inhibition of tunnelling of topological edge states by introducing out-of-phase modulations in microcavity pillars constituting a polariton topological insulator. The modulation was found to drastically affect the displacement of the states, without destroying their internal structure. Also, we showed that efficient control of the displacement is possible for modulation frequencies close to frequencies at which inhibition of tunnelling occurs in two-pillar system. Our findings are relevant to optoelectronic devices based on topologically protected edge states. The concept may be also relevant in the case of nonlinear excitations and gap solitons[59,60].

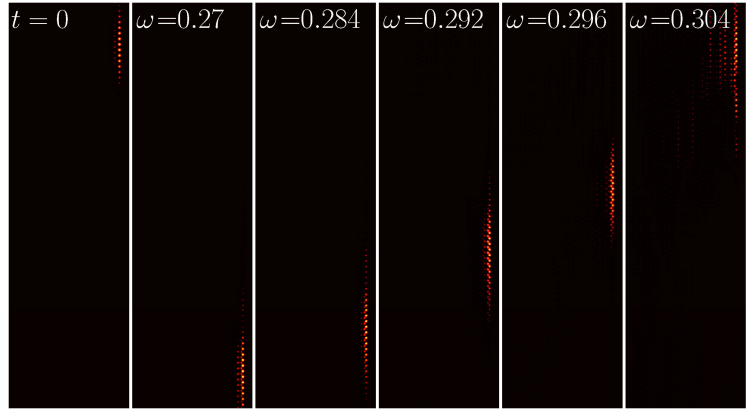

Fig. 6. Same as in Fig. 5, but for modulation frequencies around secondary resonance. Left panel shows input wavepacket. Distributions are shown within $x\in[-14,42]$ and $y\in[-90,100]$ windows. The initial edge state is launched at $y_{\rm in}=80$.

The work is supported by National Key R&D Program of China (2017YFA0303703), Natural Science Foundation (2017JZ019) and Postdoctoral Science Foundation (2017BSHTDZZ18) of Shaanxi Province, China Postdoctoral Science Foundation (2016M600777), ITMO University Visiting Professorship via the Government of Russia Grant 074-U01. Y.V.K. and L.T. acknowledge support from the Severo Ochoa Excellence Program (SEV-2015-0522), Fundacio Privada Cellex, Fundacio Privada Mir-Puig, and CERCA (Generalitat de Catalunya). Y.V.K. acknowledges funding of this study by RFBR and DFG according to the research project № 18-502-12080.

**References**

1. M. Z. Hasan and C. L. Kane, "Topological insulators," Rev. Mod. Phys. **82**, 3045-3067 (2010).
2. X.-L. Qi and S.-C. Zhang, "Topological insulators and superconductors," Rev. Mod. Phys. **83**, 1057-1110 (2011).
3. F. D. M. Haldane, "Model for a quantum Hall effect without Landau levels: condensed-matter realization of the 'parity anomaly'," Phys. Rev. Lett. **61**, 2015-2018 (1988).
4. C.-Z. Chang, J. Zhang, X. Feng, J. Shen, Z. Zhang, M. Guo, K. Li, Y. Ou, P. Wei, L.-L. Wang, Z.-Q. Ji, Y. Feng, S. Ji, X. Chen, J. Jia, X. Dai, Z. Fang, S.-C. Zhang, K. He, Y. Wang, L. Lu, X.-C. Ma, and Q.-K. Xue, "Experimental observation of the quantum anomalous Hall effect in a magnetic topological insulator," Science **340**, 167-170 (2013).
5. Z. Yang, F. Gao, X. Shi, X. Lin, Z. Gao, Y. Chong, and B. Zhang, "Topological acoustics," Phys. Rev. Lett. **114**, 114301 (2015).
6. C. He, X. Ni, H. Ge, X.-C. Sun, Y.-B. Chen, M.-H. Lu, X.-P. Liu, and Y.-F. Chen, "Acoustic topological insulator and robust one-way sound transport," Nat. Phys. **12**, 1124-1129 (2016).
7. S. D. Huber, "Topological mechanics," Nat. Phys. **12**, 621-623 (2016).
8. G. Jotzu, M. Messer, R. Desbuquois, M. Lebrat, T. Uehlinger, D. Greif, and T. Esslinger, "Experimental realization of the topological Haldane model with ultracold fermions," Nature **515**, 237-240 (2014).
9. M. Leder, C. Grossert, L. Sitta, M. Genske, A. Rosch, and M. Weitz, "Real-space imaging of a topologically protected edge state with ultracold atoms in an amplitude-chirped optical lattice," Nat. Commun. **7**, 13112 (2016).
10. M. C. Beeler, R. A. Williams, K. Jimenez-Garcia, L. J. LeBlanc, A. R. Perry, and I. B. Spielman, "The spin Hall effect in a quantum gas," Nature **498**, 201-204 (2013).

11. C. J. Kennedy, G. A. Siviloglou, H. Miyake, W. C. Burton, and W. Ketterle, "Spin-orbit coupling and quantum spin Hall effect for neutral atoms without spin flips" Phys. Rev. Lett. **111**, 225301 (2013).
12. F. D. Haldane and S. Raghu, "Possible realization of directional optical waveguides in photonic crystals with broken time-reversal symmetry," Phys. Rev. Lett. **100**, 013904 (2008).
13. Z. Wang, Y. Chong, J. D. Joannopoulos, and M. Soljacic, "Observation of unidirectional backscattering-immune topological electromagnetic states," Nature **461**, 772-775 (2009).
14. N. H. Lindner, G. Refael, and V. Galitski, "Floquet topological insulator in semiconductor quantum wells," Nat. Phys. **7**, 490-495 (2011).
15. M. Hafezi, E. A. Demler, M. D. Lukin, and J. M. Taylor, "Robust optical delay lines with topological protection," Nat. Phys. **7**, 907-912 (2011).
16. R. O. Umucalilar and I. Carusotto, "Fractional quantum Hall states of photons in an array of dissipative coupled cavities," Phys. Rev. Lett. **108**, 206809 (2012).
17. A. B. Khanikaev, S. H. Mousavi, W.-K. Tse, M. Kargarian, A. H. Mac-Donald and G. Shvets, "Photonic topological insulators," Nat. Mater. **12**, 233-239 (2012).
18. W.-J. Chen, S.-J. Jiang, X.-D. Chen, B. Zhu, L. Zhou, J.-W. Dong, and C. T. Chan, "Experimental realization of photonic topological insulator in a uniaxial metacrystal waveguide," Nat. Commun. **5**, 5782 (2014).
19. J.-W. Dong, X.-D. Chen, H. Zhu, Y. Wang, and X. Zhang, "Valley photonic crystals for control of spin and topology," Nat. Mat. **16**, 298-302 (2017).
20. M. C. Rechtsman, J. M. Zeuner, Y. Plotnik, Y. Lumer, D. Podolsky, F. Dreisow, S. Nolte, M. Segev, and A. Szameit, "Photonic Floquet topological insulators," Nature **496**,196-200 (2013).
21. L. J. Maczewsky, J. M. Zeuner, S. Nolte, and A. Szameit, "Observation of photonic anomalous Floquet topological insulators," Nat. Commun. **8**, 13756 (2017).
22. S. Mukherjee, A. Spracklen, M. Valiente, E. Andersson, P. Öhberg, N. Goldman, R. R. Thomson, "Experimental observation of anomalous topological edge modes in a slowly driven photonic lattice," Nat. Commun. **8**, 13918 (2017).
23. M. A. Bandres, M. C. Rechtsman, and M. Segev, "Topological photonic quasicrystals: Fractal topological spectrum and protected transport," Phys. Rev. X **6**, 011016 (2016).
24. S. Weimann, M. Kremer, Y. Plotnik, Y. Lumer, S. Nolte, K. G. Makris, M. Segev, M. C. Rechtsman, and A. Szameit, "Topologically protected bound states in photonic parity–time-symmetric crystals," Nat. Mater. **16**, 433-438 (2016).
25. A. V. Nalitov, D. D. Solnyshkov, and G. Malpuech, "Polariton Z topological insulator," Phys. Rev. Lett. **114**, 116401 (2015).
26. C.-E. Bardyn, T. Karzig, G. Refael, and T. C. H. Liew, "Topological polaritons and excitons in garden-variety systems," Phys. Rev. B **91**, 161413(R) (2015).
27. T. Karzig, C.-E. Bardyn, N. H. Lindner, and G. Refael, "Topological polaritons," Phys. Rev. X **5**, 031001 (2015).
28. C.-E. Bardyn, T. Karzig, G. Refael, and T. C. H. Liew, "Chiral Bogoliubov excitations in nonlinear bosonic systems," Phys. Rev. B **93**, 020502(R) (2016).
29. Y. V. Kartashov and D. V. Skryabin, "Modulational instability and solitary waves in polariton topological insulators," Optica **3**, 1228-1236 (2016).
30. C. Li, F. Ye, X. Chen, Y. V. Kartashov, A. Ferrando, L. Torner, D. V. Skryabin, "Lieb polariton topological insulators," Phys. Rev. B **97**, 081103(R) (2018).
31. Y. V. Kartashov and D. V. Skryabin, "Bistable topological insulator with exciton-polaritons," Phys. Rev. Lett. **119**, 253904 (2017).
32. P. St-Jean, V. Goblot, E. Galopin, A. Lemaître, T. Ozawa, L. Le Gratiet, I. Sagnes, J. Bloch, A. Amo, "Lasing in topological edge states of a 1D lattice," Nat. Photon. **11**, 651 (2017).
33. G. Harari, M. A. Bandres, Y. Lumer, M. C. Rechtsman,Y. D. Chong, M. Khajavikhan, D. N. Christodoulides, M. Segev, "Topological insulator laser: Theory," Science **359**, eaar4003 (2018).
34. M. A. Bandres, S. Wittek, G. Harari, M. Parto, J. Ren, M. Segev, D. N. Christodoulides, M. Khajavikhan, "Topological insulator laser: Experiments," Science **359**, eaar4005 (2018).
35. L. Pilozzi and C. Conti, "Topological cascade laser for frequency comb generation in PT-symmetric structures," Opt. Lett. **42**, 5174 (2017).

36. M. J. Ablowitz, C. W. Curtis, and Y.-P. Ma, "Linear and nonlinear traveling edge waves in optical honeycomb lattices," Phys. Rev. A **90**, 023813 (2014).
37. D. Leykam and Y. D. Chong, "Edge solitons in nonlinear-photonic topological insulators", Phys. Rev. Lett. **117**, 143901 (2016).
38. Y. Lumer, M. C. Rechtsman, Y. Plotnik, and M. Segev, "Instability of bosonic topological edge states in the presence of interactions," Phys. Rev. A **94**, 021801(R) (2016).
39. O. Bleu, D. D. Solnyshkov, and G. Malpuech, "Interacting quantum fluid in a polariton Chern insulator," Phys. Rev. B **93**, 085438 (2016).
40. D. R. Gulevich, D. Yudin, D. V. Skryabin, I. V. Iorsh, and I. A. Shelykh, "Exploring nonlinear topological states of matter with exciton-polaritons: Edge solitons in kagome lattice," Sci. Rep. **7**, 1780 (2017).
41. F. Grossmann, T. Dittrich, P. Jung, and P. Hänggi, "Coherent destruction of tunneling," Phys. Rev. Lett. 67, 516 (1991).
42. F. Grossmann, P. Jung, T. Dittrich, and P. Hänggi, "Tunneling in a periodically driven bistable system," Z. Phys. B - Cond. Mat. 84, 315 (1991).
43. M. Grifoni and P. Hänggi, "Driven quantum tunneling," Phys. Rep. 304, 229 (1998).
44. N. Goldman and J. Dalibard, "Periodically driven quantum systems: Effective Hamiltonians and engineered gauge fields," Phys. Rev. X 4, 031027 (2014).
45. G. Della Valle, M. Ornigotti, E. Cianci, V. Foglietti, P. Laporta, and S. Longhi, "Visualization of coherent destruction of tunneling in an optical double well system," Phys. Rev. Lett. 98, 263601 (2007).
46. A. Szameit, Y. V. Kartashov, F. Dreisow, M. Heinrich, T. Pertsch, S. Nolte, A. Tunnermann, V. A. Vysloukh, F. Lederer, and L. Torner, "Inhibition of light tunneling in waveguide arrays," Phys. Rev. Lett. 102, 153901 (2009).
47. I. L. Garanovich, S. Longhi, A. A. Sukhorukov, and Y. S. Kivshar, "Light propagation and localization in modulated photonic lattices and waveguides," Phys. Rep. 518, 1 (2012).
48. Y. V. Kartashov, A. Szameit, V. A. Vysloukh, and L. Torner, "Light tunneling inhibition and anisotropic diffraction engineering in two-dimensional waveguide arrays," Opt. Lett. **34**, 2906-2908 (2009).
49. M. Milicevic, T. Ozawa, P. Andreakou, I. Carusotto, T. Jacqmin, E. Galopin, A. Lemaitre, L. Le Gratiet, I. Sagnes, J. Bloch, and A. Amo, "Edge states in polariton honeycomb lattices," 2D Mat. **2**, 034012 (2015).
50. M. Milićević, T. Ozawa, G. Montambaux, I. Carusotto, E. Galopin, A. Lemaitre, L. Le Gratiet, I. Sagnes, J. Bloch, and A. Amo, "Orbital edge states in a photonic honeycomb lattice," Phys. Rev. Lett. **118**, 107403 (2017).
51. V. G. Sala, D. D. Solnyshkov, I. Carusotto, T. Jacqmin, A. Lemaitre, H. Terças, A. Nalitov, M. Abbarchi, E. Galopin, I. Sagnes, J. Bloch, G. Malpuech, and A. Amo, "Spin-orbit coupling for photons and polaritons in microstructures," Phys. Rev. X **5**, 011034 (2015).
52. S. Dufferwiel, F. Li, E. Cancellieri, L. Giriunas, A. A. P. Trichet, D. M. Whittaker, P. M. Walker, F. Fras, E. Clarke, J. M. Smith, M. S. Skolnick, D. N. Krizhanovskii, "Spin textures of exciton-polaritons in a tunable microcavity with large TE-TM splitting," Phys. Rev. Lett. **115**, 246401 (2015).
53. A. Yariv, Optical Electronics in Modern Communications (Oxford University, New York, 1997).
54. E. A. Cerda-Méndez, D. Sarkar, D. N. Krizhanovskii, S. S. Gavrilov, K. Biermann, M. S. Skolnick, and P. V. Santos, "Exciton-polariton gap solitons in two-dimensional lattices," Phys. Rev. Lett. **111**, 146401 (2013).
55. P. Dong, S. F. Preble, J. T. Robinson, S. Manipatruni, and M. Lipson, "Inducing photonic transitions between discrete modes in a silicon optical microcavity," Phys. Rev. Lett. **100**, 033904 (2008).
56. C. Schneider, K. Winkler, M. D. Fraser, M. Kamp, Y. Yamamoto, E. A. Ostrovskaya, S. Hofling, "Exciton-polariton trapping and potential landscape engineering," Rep. Prog. Phys. **80**, 016503 (2017).

57. T. Ozawa, H. M. Price, N. Goldman, O. Zilberberg, and I. Carusotto, "Synthetic dimensions in integrated photonics: From optical isolation to four-dimensional quantum Hall physics," Phys. Rev. A **93**, 043827 (2016).
58. U. Naether, Y. V. Kartashov, V. A. Vysloukh, S. Nolte, A. Tünnermann, L. Torner, and A. Szameit, "Observation of the gradual transition from one-dimensional to two-dimensional Anderson localization," Opt. Lett. **37**, 593 (2012).
59. T. Mayteevarunyoo and B. A. Malomed, "Stability limits for gap solitons in a Bose-Einstein condensate trapped in a time-modulated optical lattice," Phys. Rev. A **74**, 033616 (2006).
60. T. Mayteevarunyoo, B. A. Malomed, and M. Krairiksh, "Stability limits for two-dimensional matter-wave solitons in a time-modulated quasi-one-dimensional optical lattice," Phys. Rev. A **76**, 053612 (2007).